# 2013 Snowmass Community Summer Study White Paper

## Low Background Materials and Assay

### A Supplement to the Cosmic Frontier CF1 Summary

J. Cooley, P. Cushman, E.W. Hoppe, J.L. Orrell, R.W. Schnee

## Executive Summary

The next generation of direct detection dark matter experiments will require stringent radiopurity in their target materials, internal instrumentation, and shielding components. A survey of major material assay facilities worldwide and of anticipated assay needs of generation 2 (G2) experiments indicate the current assay capability is marginally adequate in sensitivity and inadequate in throughput. High purity germanium (HPGe) gamma-ray assay has been the workhorse for material screening. It appears hundreds of samples will require screening at the limits of current achievable HPGe sensitivity. Mass spectroscopy methods and neutron activation analysis have demonstrated superior assay sensitivity in some specific cases. However, they are also currently throughput limited. In addition, both alpha/beta screening and radon emanation analysis at or beyond the current achievable sensitivity levels will be needed for dozens of samples.

Experiments that follow G2 will have even more stringent material radiopurity specifications, exceeding current assay capabilities in both sensitivity and throughput. These screening needs must precede the commissioning of experiments by 3-5 years to inform design and quality control of components. Achieving the radiopurity goals of the next decade's experiments will require investment in new techniques and tools to improve material assay sensitivity and throughput. Engineered low-background materials are a significant R&D expenditure, requiring developmental lead-time, and should benefit many end-user experiments. Reserving underground real estate for assay methods affected by cosmic rays (e.g. HPGe) and engineered radiopure materials activated by cosmic rays (e.g. electroformed copper) is judicious forward planning. In many cases, these spaces will require radon-suppressed sample preparation and material storage space.

The above-mentioned survey of material assay facilities indicated very little development is taking place to improve sensitivity or throughput capability. In order to create the necessary infrastructure, it may help to form a consortium of low-background assay centers in the U.S that is managed by a scientific board with representatives from the community. This would provide for common use of existing screening facilities and a unified plan for new infrastructure and site development (see the Facilities White Paper). Such a loose organization already exists as AARM (Assay and Acquisition of Radiopure Materials), which is currently populating a community-wide materials database with published assay information. A more formal organization such as a Consortium is being actively investigated with DOE and NSF. Eventually, results of all screening performed in Consortium assay centers could be input directly to the open-access database, reducing duplication and providing for more efficient vendor selection.

## I. Low background experiments – The challenge

Next generation experiments in dark matter, neutrinoless double beta decay, and low-energy neutrino detection will require special attention to background reduction methods to achieve their research goals. Most of these experiments are performed underground to reduce external backgrounds induced by cosmic rays. The scientific research areas of these experiments span the traditional disciplines of nuclear, particle, and astroparticle physics. As the reach of dark matter (lower cross section sensitivity), neutrinoless double beta decay (longer half-life sensitivity), and low-energy neutrino detection (rate and spectral-shape sensitivity) has grown, experiments have increased in size and demanded lower background rates to achieve their science goals. As described below, background reduction falls into two broad categories: experimental design ingenuity and direct management of the background source terms. A review of contemporary published experimental results from these fields shows that for nearly all experiments, the limiting background is due to source terms that are internal to the experimental construction materials. While further innovation in technological design may make future experiments more insensitive to, better able to discriminate against, or able to avoid the impact due to internal backgrounds, this whitepaper addresses the issue of background mitigation by directly managing the source terms of the background in question. As the vast majority of internal backgrounds are due the presence of naturally occurring primordial radioactive isotopes (principally those in the uranium decay chain, thorium decay chain, and potassium), directly managing the background source term encompasses addressing the *materials* containing these problematic isotopes and using some means of *assay* to ensure during the construction process of large scale experiments the intended background goals are met and the scientific reach is achieved. This approach to background reduction is referred in short hand as "low background materials and assay."

## I.1 Methods of background reduction

There are several methods for addressing backgrounds in experimental detector systems:

- Insensitivity to a potential background
    - COUPP's gamma-ray insensitivity
- Discrimination (against) the background
    - SuperCDMS's nuclear recoil vs. electron recoil discrimination
- Fiducial volume cuts
    - LUX's self-shielded inner volume and position reconstruction
- Characterization of the background
    - *In situ* and *ex situ* assay to robustly quantify the background rates
- Material handling
    - Cleanrooms; DarkSide's reduced radon environments
- Material selection
    - EXO's material assay tome: NIMA 591 (2008) 490–509
- Material purification
    - Majorana Demonstrator's electroformed copper to reduce U/Th isotopes

Generally, the top three methods are the basis for the design of a specific experiment and represent the physicist's tool kit for implementing a scientific research program. Background characterization can take place either by using the instrument itself (*in situ* background characterization) or by some other assay means performed outside of the experimental instrument (*ex situ* background characterization). In the former case a good experimental design is crucial so that *in situ* characterization can address its own uncertainties in measurement. In the latter case, an *ex situ* measurement of the source term of some background is made and then a simulation or other method for inferring a specific background level from the *ex situ* characterization is required. The last three background mitigation methods can either be developed within an experimental program for a specific target measurement or in other cases rely upon pre-existing resources and experts in material characterization such as spectroscopists, analytical chemists, or material scientists depending on the specific measurement or material need.

## I.2 Backgrounds reported in the contemporary literature

A short review of recently reported results and background predictions is contained within Appendix A. The vast majority of backgrounds reported are due to primordial radioactivity present in the materials used in the construction of the experiment. That is, the source of the background is generally consistent across experiments. Answering *why* the primordial radioactivity creates a background varies by experiment. In some cases experiments can discriminate against some types of backgrounds, but the discrimination may not be perfect and thus a "leakage" of background events may occur. In other cases the experiment cannot discriminate the background and thus the events simply appear in the data set and must be estimated. Lastly, in some cases, the handling of the detector construction materials is insufficiently clean, resulting in elevated background rates from the expected values. Regardless of the *why*, the source is still nearly always primordial radioactivity. Methods that reduce or quantify the levels of these primordial background sources will directly contribute to mitigating their impact in future experiments.

## II. Radiometric Assay technology and sensitivity

The type of material assay required depends on the target isotope or impurity, the material to be screened, and the required level of sensitivity. Therefore, a variety of techniques and facilities will be necessary, which must be matched to the needs of each experiment. This section presents the most common methods for radioactive impurity screening of materials using radiometric techniques that measure the radiation emitted from the material under study.

## II.1   High purity germanium (HPGe) gamma-ray assay

High purity germanium (HPGe) gamma ray spectroscopy is a mature technology that is seen as the primary tool for material assay as part of material screening and selection programs. For this reason, the use of HPGe detectors for assay screening of materials is given a longer description.

When applied to assay for the uranium decay chain, HPGe assay is principally sensitive to the gamma rays from $^{214}$Pb and $^{214}$Bi at the bottom of the decay chain and $^{226}$Ra (though significantly less intense) located in the middle of the decay chain, immediately above $^{220}$Rn. The noble gas $^{220}$Rn is important as it has a 3.8-day half-life and can potentially migrate out of the matrix material creating disequilibrium in the uranium decay chain within the material being assayed. This may or may not be an issue; it depends on the target impurity isotope that is a concern as a background source for a given experiment. For example, if the concern is spontaneous fission neutrons from $^{238}$U, then assay via sensitivity to $^{214}$Pb and $^{214}$Bi may provide misleading results if the equilibrium of the decay chain is disrupted (e.g. during processing or due to $^{220}$Rn migration). Radium participates in aqueous chemistry and can similarly lead to broken chain equilibrium at $^{226}$Ra during the industrial production of materials. However, if the experiment's background is gamma rays from $^{214}$Bi, for example, then a direct measurement of the gamma rays from $^{214}$Bi is usually ideal.

When applied to assay for the thorium decay chain, HPGe assay is principally sensitive to the gamma rays from $^{208}$Tl, $^{212}$Bi, and $^{212}$Pb located at the bottom of the decay chain and $^{228}$Ac located toward the top of the decay chain above $^{220}$Rn (only a 56 second half-life). In many cases the background of concern for experiments is precisely the high-energy 2615 keV gamma ray from $^{208}$Tl that conspicuously dominate an HPGe assay spectrum from the equilibrium thorium decay chain. In these cases HPGe gamma-ray assay is a likely method for assay material screening.

In addition to gamma radiation emission, there are many alpha decays in both the uranium and thorium decay chains. If the background for an experiment is directly alpha particles or secondarily neutrons from induced (α,n) reactions, HPGe screening may or may not be the assay of choice depending on an experiment's level of sensitivity to these alpha-related backgrounds and their location.

See Appendix D for the primary U/TH decay chains with noted gamma-assay isotopes.

The decay of $^{40}$K produces a 1461 keV gamma ray. If an experiment is concerned with backgrounds from $^{40}$K gamma rays, HPGe screening for $^{40}$K is typically the correct approach.

HPGe detector systems used for material screening take many different shapes, sizes, designs, and locations. Three general categories of HPGe instruments are described.

*Commercial systems*

HPGe detectors are sold commercially by a number of detector vendors. From the perspective of material assay screening, the primary specification of interest is the detector's relative efficiency compared to a 3"x3" NaI detector [IEEE]. The relative efficiency is directly related to the counting efficiency of the detector system. HPGe detectors range from 20%-150% relative efficiency with costs ranging from $20k to $100k+, depending on the additional features requested. If the HPGe detector is being used as a screening counter, the commercial vendors can provide a number of additional features (at additional cost) including single-unit lead shielding, spectroscopic measurement systems, and software to analyze gamma-ray spectra to identify isotopes via their gamma-ray emission. Such systems are best suited for rapid screening to test for levels of radioactivity that are at the level of naturally occurring U/Th/K found in soils etc. Testing for U/Th/K levels in refined and purified materials typically requires more sensitive systems. However, in some cases the radioactivity-screening requirement for a given material is not particularly stringent (e.g., outer shielding materials) and a confirmatory screening reporting the activity is "less than the experimental requirement" is sufficient. Identifying these cases when less stringent material screening requirements are needed is an important part of material assay screening management: if the sample doesn't need a sensitive measurement, don't put it on the most sensitive HPGe counter to preserve that resource for more sensitive screening.

*Augmented commercial systems*

An augmented commercial system is typically a commercial HPGe germanium detector that the owner (purchaser) has placed inside a specially designed ("custom") shield, perhaps including graded shielding materials (such as lead and copper), use of neutron moderation and capture materials, a radon mitigation system, an active cosmic ray veto shield, and even an underground location. However, the germanium detector itself remains in the original commercial cryostat, often in a vendor's low-background-materials version of their standard cryostats. In these cases the cryostat and shielding materials will typically determine the sensitivity achieved by the screening detector. As an example of this type of detector system, an HPGe material assay screening system located deep underground at SNOLAB reported the following sensitivity levels [Lawson]:

| Isotope/Chain | Standard Size (ppb) \| (mBq/kg) | Large size & Long count (ppb) | Typical for Earth's Curst (ppm) \| (Bq/kg) |
|---|---|---|---|
| U-238 | ~0.1 \| ~1.0 | 0.009 | 3 \| 37 |
| Th-232 | ~0.3 \| ~1.5 | 0.02 | 11 \| 45 |
| K-40 | ~700 \| ~21 | 87 | [2.5%] \| 800 |

*Fully custom systems*

A fully custom HPGe screening detector typically tries to take advantage of all the shielding augmentations listed in the augmented commercial system category with the addition of special cryostat design and attention to design and placement of the electronics read-out components. These efforts are implemented to drive down the background sources (U/Th/K etc.) near the HPGe detector. Choice of low-background shielding materials (e.g. old or ancient lead, high-purity copper) is also a typical design feature for these systems. When successful, these are truly state-of-the-art, world-class sensitivity systems. Probably the world's most sensitive HPGe material assay screening system is the GeMPI detector operated underground at LNGS [HLN]. The reported sensitivity of this system is:

| Isotope/Chain | Best sensitivity (long count) (ppb) \| (mBq/kg) |
|---|---|
| U-238 | 0.001 \| 0.012 |
| Th-232 | 0.001 \| 0.004 |
| K-40 | 1 \| 0.031 |

In reference to fully custom systems having overburden shielding from being located underground, based on an "Overview of the screening activities with HPGe detectors" [Laub] in Europe, one can infer at what laboratory depth (overburden) the cosmic ray muon flux is sufficiently shielded such that the materials of the HPGe detector system limit the sensitivity of the system. In the units of normalized counting rate (total counts in a range 300-3000 keV divided by the count time), systems both in shallow and deep underground labs reach a sensitivity limit around $10^2$ counts/day/kg. The figures below provide this information. The plot on the left has been updated by G. Heusser [Heu] to include improvements to an HPGe material assay screening detector located in a 15 m.w.e. lab at MPI-K-HD that reaches below $10^3$ counts/day/kg.

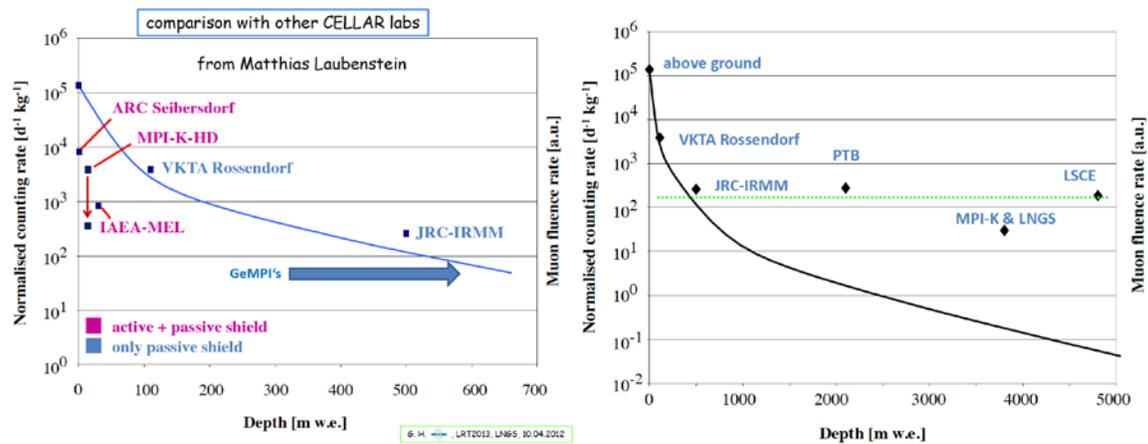

The dividing lines between the above categories are not clear-cut since a true continuous spectrum of instrument design and sensitivity exists. Other augmentations are possible as well, such as thin windows and low thresholds that are useful for extending background characterization into the x-ray region (useful for solar neutrino projects) and potentially understanding cosmogenic activation. Thus, the above discussion is intended to provide a general basis for understanding the nature of the HPGe detector as a screening instrument.

One difficulty with ultra-high sensitivity HPGe gamma-ray assay measurements is the long duration required to acquire the statistics for a significant conclusion at the ultimate sensitivity level of the instrument. For example, 24 kg of OFHC copper nuggets were counted on the GeMPI-2 detector at Gran Sasso for 4 months to determine the U/Th/K levels at 20/30/190 microBq/kg levels.

In contrast to most of the other techniques described in this document, HPGe detectors provide non-destructive assay. This is especially useful for screening the actual – often high-value – instrumentation that will be put into a low background experiment. In a similar vein, HPGe gamma-ray counting can assay complex instrumentation composed of numerous materials (e.g., PMTs composed of glass, metal, and plastic) that may not be readily chemically processed into another instrument standard sample type.

Appendix C presents a more complete (though still not exhaustive) list of HPGe screening detectors around the world.

## II.2 Alpha/Beta Counting

There is also a need for alpha and beta screening for contaminants that are not accompanied by gamma emission. $^{210}$Pb and its progeny do not have a penetrating signature and are deposited on all surfaces exposed to radon. Such radon plate-out plagues all rare-events searches, since it creates patina of contaminant that causes nuclear recoils, beta-emission, and alphas. In addition, since Pb is often used in circuitry, alpha-decay of the $^{210}$Pb progeny $^{210}$Po can cause single-site upsets. Surface contaminants, such as $^{40}$K, and anthropogenic contaminants like $^{125}$Sb and $^{137}$Cs are also detectable by beta screening.

Typical "industry standard" alpha spectroscopy is available in NIM-rack mounted instruments. The preparation of the alpha sample almost uniformly requires quantifiable analytic chemical dissolution and sample preparation to create a sample that will allow the alpha particle emission. These standard techniques are directed toward measuring material's bulk alpha activity and are clearly a destructive assay method. However, the dissolution (potentially additional chemical separations) can allow for highly sensitive measurements with the remaining alpha emitting isotopes, at levels largely dependent on the chemical preparation process that concentrates the alpha emitting elements.

The XIA Ultralo-1800 alpha particle counter [XIAa] is an ionization counter with a sensitivity of 0.0011 +- 0.0003 alphas/cm$^2$/hr. The XIA has a drift chamber 15 x 21 x 21 inches in size which is filled with boil-off gas from a liquid argon dewar. The counting region is adjustable to a square, 1800 cm$^2$ region or a circular, 707 cm$^2$ region, allowing non-destructive screening of large surfaces. Unlike proportional counters, the XIA counter is able to distinguish alpha particles originating on the sample tray from alpha particles originating on the chamber walls, ceiling or mid-air using pulse shape analysis [XIAb].

Improved sensitivity to *surface* alpha contamination should be achievable using a multi-wire drift chamber, which would provide a much lower surface area of detector materials in the fiducial volume and improved rejection based on tracking of decays not from the sample. By using a radiopure gas such as neon, clean materials for the detector construction, and passive shielding similar to that used for HPGe detectors, such a detector may also be directly sensitive to surface beta decays, providing sensitivity much better than typical HPGe detectors to $^{210}$Pb on detector surfaces. The primary challenge, of construction of radiopure multi-wire proportional counters with uniform wire gain, sufficiently low outgassing, and small noise, has been successfully demonstrated [8].

## II.3 Proportional Counters

Gaseous proportional counters have not typically been used for material assay characterization of materials for low background experiments. However, because they can be sensitive to charged particles emitted from surfaces, it is potentially feasible to employ proportional counter techniques to screen surface emission, although developing an appropriate sample introduction mechanism is challenging.

A notable exception to this is the case of measuring the presence of (naturally occurring) $^{39}$Ar in geologically aged argon from underground wells. The argon gas from underground wells is depleted in $^{39}$Ar relative to atmospherically sourced argon gas. Since $^{39}$Ar is a beta-emitter, the geologic argon is of interest as a potential low-background detection medium and/or active veto. However, it has been demonstrated [Xu] that liquefied argon self-counting detectors located underground can readily achieve sensitivity levels well below what is achieved with the lower mass (volume) of gas that can be analyzed in gaseous proportional counters.

## II.4 Radon emanation chambers

Often, the dominant radon-induced background in an experiment is due to emanation of radon from the detector materials [Simgen]. Direct measurement of the rate of radon emanation from detector materials is therefore important. Such emanation measurements may in some cases also provide more sensitive determination of the Ra content of materials (0.1-10 µBq/kg) than is typically achievable with HPGe screening. Samples are isolated in a vacuum chamber, either at vacuum or with a carrier gas (typically $N_2$ or He). Pumping the emanated atoms (with or without the carrier gas) through one or more traps may be done with sufficiently high efficiency that sensitivities to <10 atoms have been achieved, very close to the ultimate possible. Detection may be done most efficiently either with ultra-low background Lucas cells [Liu] or miniaturized proportional counters [Rau, ZS], though electrostatic detectors are more common in practice. For most of these set-ups, the transfer time from chamber to detector is long enough that the method is strictly limited to $^{222}$Rn, but some setups can provide fast enough transport that $^{220}$Rn and even $^{219}$Rn may be measured.

## II.5 Immersion Whole Body Counting

While an augmented suite of sensitive production screeners can provide the bulk of the assay, orders of magnitude improvement in sensitivity is required for some materials close to detectors and for active elements. An ultra-sensitive whole body screener in a water tank at depth would provide an ultimate check on the total activity from all isotopes in the material, including short-lived isotopes that are impossible to detect chemically. The goal of such a system is bulk assay of large amounts of material at the $10^{-13}$-$10^{-14}$ g/g U/Th level. Such designs have been explored in NUSL [Nico] and DUSEL reports [Cushman] and generally resemble the Borexino Counting Test Facility (CTF) [Ali] with a top-loading sample changer.

## III. Non-radiometric Assay Techniques

For many applications, low-level counting is not necessarily the best technique for screening materials, particularly in those cases where the radioisotope is of low specific activity. Radioisotope identification can be done using surface techniques and mass spectrometry. Most research universities and many companies offer such services for a fee. University fees are often relatively inexpensive for faculty, and companies offer quicker turn-around for a higher charge. National Laboratories often have state of the art equipment in niche areas developed for science outside of the field of physics. To gain the most benefit for the community without duplication, it may be useful to forge user agreements with existing labs or encourage commercial analytical labs to open branches to support underground science. A summary of the more common techniques available follows.

### III.1 Mass Spectroscopy

This suite of techniques extracts and accelerates charged ions from the sample, separates them according to atomic mass in a magnetic or electrical field, and then measures the current of ions with a detector (often a Faraday cup) which intersects the trajectory corresponding to the correct charge-to-mass ratio for the element in question. The sensitivity thus depends on both the resolution of the mass spectrometer, as well as the efficiency associated with the sample dispersion technique by which atoms or ions are introduced and accelerated into the channel.

Inductively-Coupled Plasma and Thermal Ionization (ICPMS, TIMS) must first put the sample into solution using various combinations of concentrated acids or bases. In TIMS a filament is coated with the solution and then heated, whereas for ICPMS, a flow of gas (usually argon) converts the liquid sample into a fine aerosol. A portion of the sample aerosol is then directed through the center of an argon plasma torch, where the atoms are ionized. All samples must be background subtracted using a blank (the dilute solution containing the preparation reagents without sample) and normalized to a standard (pure trace element in same solution) since the ionization efficiency depends on the element. Instruments capable of probing to ppb levels are typically found in geology and chemistry departments in many universities and also in many commercial analytical labs. In order to break the ppb barrier requires clean room conditions to avoid contamination of samples, blanks, and standards for any of the mass spectrometry methods. Also elemental isolation or chemical purification techniques to avoid the isobaric interferences caused by other isotopes or ion complexes with the same mass is often required when large quantities dissolved solids are present in solutions. With these efforts ppt to ppq sensitivities are realized, often beyond what radiometric counting techniques can achieve.

Secondary Ion and Glow Discharge (SIMS, GDMS) release sample atoms via surface bombardment. In the case of SIMS, an ion beam sputters the surface and the released ions enter the spectrometer, whereas for GDMS the bombarding ions come from a low pressure DC plasma discharge cell in which the sample is the cathode. The sputtered neutral atoms are then ionized in the plasma and extracted into the mass analyzer for separation and detection, whereas sputtered positive ions simply fall back onto the sample and don't make it into the spectrometer. Thus GDMS is not as matrix-dependent as SIMS.  Both techniques can be conducted directly on samples with little or no preparation or separation chemistry and offer advantages of quick turnaround compared to other mass spectrometric analysis methods requiring time- and labor-intensive preparation.  Both offer the ability to form a depth profile as the sputter proceeds although SIMS is capable of microscopic physical spatial resolution.  Conductive materials are easiest to analyze by GDMS, though electrodes can be formed.  SIMS is sensitive to the ppb levels while GDMS is excellent for identifying trace elements in bulk samples down to tens to hundreds of ppt.

Accelerator Mass Spectroscopy (AMS) is done at a number of centers around the country, typically where tandem accelerators used for nuclear research have been retooled for this application.  After an initial spectrometer step to remove background, the ions are accelerated to MeV energies and passed through a second magnetic analyzer, where detection is typically by ion chambers and multi-wire ion counters, achieving ppt and even sub-ppt levels.  Since this sensitivity can be achieved in a few hours on samples smaller than 1 mg (where simply counting the decays would take several human lifetimes), AMS has become the technique of choice for $^{14}$C dating.  It is expensive and not readily available as a "consumer" instrument; instead, a number of labs have sprung up to prepare and isolate samples (e.g. from cores [Nico]) and then take care of sending the samples off and keeping track of the results

Radioisotope identification using mass spectrometry does not require shielding and can be performed on the surface.  While instruments capable of probing to ~$10^{-9}$ g/g levels are typically found in Universities and commercial analytical laboratories, the ~$10^{-12}$ to $10^{-15}$ g /g levels achieved by inductively-coupled plasma mass spectroscopy (ICPMS) are only realized at a few locations where radiochemical expertise and careful attention to quality assurance protocols are augmented by novel dissolution and digestion techniques which can process a wide variety of sample types.  Since ICPMS provides an alternative to counting, it must be taken into account in any determination of future assay needs, as well as included in any centralized scheduling apparatus. In fact, at the assay levels required for future experiments it is likely to become the primary assay technique due to its higher sensitivity particularly for low activity radionuclides.

### III.2 Neutron Activation Analysis

Even though technically a decay-counting technique, the enhanced signatures generated by Neutron Activation Analysis (NAA) means that this technique does not need a shielded underground site to achieve ppt sensitivity.  A source of neutrons is required to initiate a neutron capture interaction in the sample.  This source is generally a reactor with fluxes of $10^{13}$ neutrons cm$^{-2}$ s$^{-1}$, but new deuterium-tritium plasma generators are approaching this intensity [Reij]. The resulting compound nucleus forms in an excited state, almost instantaneously de-exciting into a more stable configuration through emission of one or more characteristic prompt gamma rays. In many cases, this new configuration yields a radioactive nucleus that also decays (often via beta-decay) followed by emission of one or more characteristic gamma rays, but at a much slower rate according to the unique half-life of the radioactive nucleus.  Observing the prompt gammas during irradiation is called PGNAA, but the more common procedure is to remove the sample from the reactor and observe the characteristic gammas for the longer lived isotopes formed, usually via high purity germanium detectors. Since shipment of samples irradiated at US research reactors can be arranged in a time scale of 2 to 3 days, short-lived products are most efficiently counted at the irradiation site. For many longer-lived activation products gamma screeners can be used. A typical exercise would be counting $^{239}$Np ($t_{1/2}$=2.36 d) and $^{233}$Pa ($t_{1/2}$=27 d) from $^{238}$U and $^{232}$Th.

This technique is limited by the nuclear properties of the trace element and the substrate that it is contaminating. Approximately 30% of the elements do not have reactions that can be probed in this way and for those that do, the activation of the substrate can sometimes mask the lines of interest used for analysis. Detection limits assuming interference-free spectra range from 1 pg of Eu to 100 pg of U to 10 ng of K. [Parry, Gold]. To reach the ppt range or below, the elements of interest are typically chemically separated from interfering side activities [Djurcic].

Materials containing light metals are typically not suitable for NAA and even many polymers can contain sufficient quantities of sodium and other contaminants to render NAA ineffective.

Typically NAA is considered a destructive assay technique, as it is unlikely the irradiated material may then later be used in the experimental apparatus. Thus, the use of NAA applies to measurements of samples from lots of materials to provide some additional confidence that the sample measured in NAA is representative of the materials actually employed in the experimental construction.

### III.3 Surface analysis

Ion Beam Analysis (IBA) [Jeynes, Comp] utilizes high-energy ion beams to probe elemental composition non-destructively as a function of depth to several microns with a typical depth resolution of 100-200 angstroms. The energy distribution of backscattering ions (RBS = Rutherford Backscattering Spectroscopy) quantifies the depth distribution for a given element. Distinctive characteristic X-rays emitted from the different target elements (PIXE = Particle-Induced X-Ray Emission) upon beam bombardment ensure the accurate identification of similar mass elements. Additional simultaneous measurements include forward-recoil spectrometry (FReS) or a NaI scintillator detector for Nuclear Reaction Analysis (NRA). These techniques can identify a monolayer of surface contamination. Electron beams for Auger Electron Spectroscopy are less sensitive; the best can find trace contaminants at the 0.1% level within 1 nm of the surface. However, it is possible to obtain depth profiles of 100 nm or more with a sputter gun. Spot size can be small enough (5 microns) to do an x-y probe of contaminants on the surface.

### III.4 Atom Trap Trace Analysis (ATTA)

Laser cooling techniques can be used to trap atoms, excite them to a metastable state, and then detect their fluorescence, thus determining abundances by directly counting atoms [Chen]. Atom Trap Trace Analysis (ATTA) provides a fast turn-around method of measuring radioactive background from $^{85}$Kr and $^{39}$Ar to a few parts in $10^{-14}$ [Collon] and could be installed underground to screen user samples, as well as aid in the purification of Ar, Ne, and Xe.

## IV. Assay Needs

Determining the number and sensitivity of the assay capabilities needed by the field must take account of the long lead time required for low-level assay and the screening needs preceding the commissioning of each experiment, typically by 3-5 years, as the assays inform design, as well as establish quality control of all pre-installed components.

| Technique | > 1 mBq/kg | 0.05 – 1 mBq/kg | < 0.05 mBq/kg |
|---|---|---|---|
| HPGe Gamma Screening | 160 | 400 | 65 |
| NAA and/or ICPMS | 95 | 385 | 40 |
| Radon Emanation and/or alpha/beta screening | 0 | 160 | 10 |

Table 1:  Summary of responses to the 2013 SNOWMASS on the Mississippi screening needs survey.

As part of the *2013 Snowmass on the Mississippi* Community Planning process a survey of screening needs was sent to collaborations planning direct detection dark matter experiments contemporary with the second-generation (G2) experiments. We received responses from COUPP, C-4, DarkSide, DM-Ice, LUX, SuperCDMS and XENON-1T. From the information collected, we estimate that there will be a need for ~1000 samples to undergo gamma screening in HPGe screeners and ~385 samples undergo NAA and/or ICPMS just to support those experiments responding.  In addition, we estimate that ~170 samples will need to undergo radon emanation analysis.

## V. Assay Facilities

A separate Snowmass working group is developing a white paper related to facility requirements for next generation experiments. The first subsection (V.1) presents much of the same material provided to that working group for inclusion in the Facilities and Infrastructure White Paper regarding underground facilities for radiometric assay. The material is also included here as it explains the resources required to address assay needs. The last two sub-sections (V.2 and V.3) provide additional information on facility needs for non-radiometric assay and the facility related issue of creating low-radon environments.

As a side note on underground facilities: Long-term underground storage can also mitigate the effects of surface exposure by allowing the cosmogenic activation products to decay away prior to material use. Thus, space for long-term storage of materials that can become activated by surface exposure to cosmic ray neutrons, such as copper and germanium, is also a beneficial facility use.

### V.1 Underground Facilities for Radiometric Assay

A key issue for radiometric screening is the interference of cosmic rays in the radiation measurement process. This is especially true for HPGe assay detectors, as already discussed in some detail in Section II.1. While mass spectrometry and atom trap techniques require no shielding from cosmic rays, direct radiometric gamma and the most sensitive alpha and beta counting techniques require shielding to obtain the required sensitivity, often this means measurements are performed in underground facilities.

The US today has a decentralized set of radiometric assay detectors dedicated to specific research projects at surface sites, a few shallow depth sites which have evolved into user facilities, and a few deep sites which have limited through-put to provide the ultra-low background counting (primarily gamma assay with germanium spectrometers) typically desired for screening materials for the next generation of double-beta decay, solar neutrino, and dark matter experiments. Interestingly, these ultra-sensitive screening detectors can also be used by geology, microbiology, environmental science, and national security applications to identify radioisotopes, date samples, and measure tracers introduced into hydrological or biological systems. There has not, to date, been much dialogue with other fields that might also benefit from the ultra-low counting sensitivities available at deep sites. While outreach to other research fields is recognized as a means to enhance the user base and increase capability, it has only been possible to achieve in large national laboratories with historical ties to other communities, notably homeland security applications.

Although the background is reduced significantly by moving screening detectors underground, it is not possible to exploit extreme depths, since backgrounds internal to the detectors themselves can become the limiting factor beyond an overburden of 1000 meters water equivalent. Moreover, for many practical applications, the counting time itself would become prohibitively long. Thus, an underground screening facility does not have to be as deep as the experiments for which it is screening materials, even accounting for improvements in the purification of germanium crystals and new low background construction or shielding materials. The use of simple muon veto systems can make even shallow sites such as ~15-30 meter water equivalent shallow labs sensitive enough for many applications as previously described in section II.1 on HPGe assay counters.

Existing facilities for underground radiometric assay are presented in Table 1 found in Appendix C.

In some countries the assay infrastructure is much more developed. It has been built up in conjunction with centralized laboratories such as SNOLAB in Canada and LNGS in Italy, and thus has the advantage of co-location with many of the experiments it services. The existence of centers for assay creates the critical mass of experts and organization to extend the services to other fields, producing in turn, a broad user community capable of producing a self-sustaining business model. Table 2 in Appendix C provides information on international assay infrastructure.

The original vision for DUSEL, the Deep Underground Science and Engineering Laboratory considered for location at Homestake, included the same attention to assay infrastructure as is manifested in the investments of Europe, Canada, and Japan (and now China). However, without such a national underground laboratory to concentrate screening in one location and to justify the expense, we may require a new model. Over the last decade, the gap between the needs of rare event physics searches and available screening has widened considerably, since it has been left to individual experiments to propose, individually, capabilities that are manifestly needed by the entire community. This has led to a shortage of screening infrastructure overall and uneven distribution of existing resources.

Four deep sites in the US have made infrastructure investments leading to potential low-background counting centers. The Waste Isolation Pilot Plant (WIPP) in Carlsbad, NM is at a depth of 1600 mwe within a layer of bedded, impermeable salt. WIPP is a DOE facility with a fully developed infrastructure and a heavy Los Alamos National Lab presence with an Earth and Environmental Science office located at Carlsbad. Although fundamental science is only an add-on to its primary mission, LANL and WIPP established a clean room to house the Majorana collaboration's segmented multi-element germanium arrays (SEGA and MEGA), and is also adding facilities to service the EXO double beta decay experiment.

A second site is the Kimballton mine operated by the Chemical Lime Company in Giles County, Virginia. Virginia Tech is the lead institution, also developing this site for the LENS solar neutrino experiment. Kimballton's proximity to the Naval Research Labs make it a promising site for secure applications involving homeland security and treaty verification; plans are already underway to make this a reality.

The third deep site is the Soudan iron mine (2090 mwe) in northern Minnesota, which is home to the MINOS, SuperCDMS and CoGeNT experiments. Currently one HPGe is being used to count samples for XENON and Majorana collaborations as part of the SOLO facility and a second is dedicated to SuperCDMS screening. The proportional tube panels that line the Soudan2 proton decay hall have been repaired and a new DAQ installed, creating a 14 x 12 x 33 $m^3$ space with time-stamped muon tracks which can be used as an offline veto or for muon studies underground. The muon tracks are correlated with two neutron detectors, a liquid scintillator detector and a gadolinium-doped neutron capture detector sitting on a large lead target to access the high energy cosmogenically-produced neutrons. This work is part of an effort to benchmark cosmogenic simulations and improve the physics modules in Geant4 and FLUKA codes.

The fourth site is the Sanford Underground Research Facility located in Lead, SD. SURF is hosting the MAJORANA DEMONSTRATOR and LUX experiments at the 4850' level (4300 mwe). The CUBED Collaboration has been developing a low background HPGe counter on the same mine level for use in screening. The former Homestake mine in which SURF now exists could potentially be expanded for future underground research and development.

Low-level counting that can be done in shielded environments on the surface is usually performed by a high-purity germanium detector (HPGe) or NaI crystals equipped with a PMT. These are commercial devices that are usually dedicated to a specific project. All the national labs have such detectors and many will negotiate a "use for others" contract with outside users. Many university groups also have dedicated machines. The great need here is for coordination and integration of these facilities. LBNL has maintained a surface site (bldg 72) surrounded by low activity concrete, currently housing NaI counters and a 130% HPGe detector, while also running an 80% p-type HPGe at the nearby Oroville Dam (180 mwe), although this facility is currently being dismantled. PNNL operates a shielded facility (~30 mwe) on its site that opened in 2010 and at present includes a 14-crystal HPGe array, a low background proportional counter array, and two commercial low-background HPGe systems with enhanced shielding. These systems use active anti-cosmic veto systems to improve their performance as cosmic ray muon are still present at these relatively shallow depths.

## V.2 Facilities for Non-radiometric Assay

In the cases of non-radiometric assay methods, the facilities required to make the most sensitive measurements are not driven by being underground, but rather relate to having a controlled environment that dramatically limit the possibility of contamination of the sample during the sample preparation process. Thus the most sensitive assay measurements are performed in cleanrooms. The use of a cleanroom and cleanroom protocols is to protect the sample preparation from contamination either by the personnel performing the work or by uncontrolled particulate in the air.

Furthermore, although a clean and controlled facility is a requirement for the most sensitive non-radiometric (analytic) measurement techniques, it is often the case that the materials used in the sample preparation process become the leading contributor to the signature of interest. For example, if mass spectrometry is used for measurement of $^{238}$U, the presence of $^{238}$U in ALL other materials that hold or process the sample prior to mass spectroscopic analysis will impact the ultimate level of sensitivity achievable.

Having a clean and controlled laboratory with preparation materials that add minimal contribution to the measurement targets also relies upon trained expert technicians able to perform sample preparation and final measurements routinely without negatively impacting the instrument.

## V.3 Facilities for controlling radon

Of special concern is the impact of radon. It is difficult to create radon-free environments as radon noble gas can pass through many materials and is emanated from materials as a result of uranium decay chain isotopes.

Radon and its daughters are an important background for all underground physics rare-event searches such as for WIMP dark matter or neutrinoless double-beta decay. Radon is present in all air, produced by radioactive decays in the uranium and thorium chains, and naturally has especially high concentrations deep underground. Gamma decays from radon's short-lived daughters may provide backgrounds, and radon daughters deposited from the atmosphere onto detector surfaces provide particularly dangerous backgrounds. Low-energy beta decays on detector surfaces or in the bulk, the $^{206}$Pb recoil nucleus from $^{210}$Po alpha decay, and neutrons from (alpha,n) reactions (especially on the fluorine in Teflon) all may produce significant backgrounds for future experiments. In order to reduce radon-induced backgrounds, the community needs to develop improved methods to reduce and sense the radon concentration in air and in detector gases and liquids.

The standard method for reducing the radon concentration of breathable air below that of outside air is to flow the air continuously through a column containing an adsorbent (typically cooled, activated charcoal at -40 °C to -70 °C) so that most radon decays before it exits the filter. Continuous systems are relatively simple and robust, are available commercially, and are demonstrated to achieve 1-10 mBq/m$^3$ [Nac, Gal]. Alternatively, it may be possible to develop improved radon reduction at a lower cost by using a swing system, where one stops gas flow well before the breakthrough time, and regenerates the filter column while switching flow to a second column. For an ideal column, no radon reaches the output. However, swing systems are more complicated than continuous systems (both in terms of their analysis and operation). Regeneration may be achieved by increasing the pressure from vacuum, or by increasing the temperature [Gra], with the latter expected to provide the best performance, albeit at higher cost and complexity than a vacuum-swing system. The best performance achieved with full-scale swing systems is in the range of 1 Bq/m$^3$ [Poc, Sch].

For most gases used in WIMP-search experiments, continuous systems at lower temperatures (<80 K) convenient when liquifying produce significantly better radon reduction, to less than or equal to one micro-Bq/m$^3$ [Heu, SZ]. The important exception is xenon, since it cannot be cooled below 170 K without freezing, and its similarity to radon makes radon's filtration from it difficult. Identification or development of materials (from zeolites to carbon molecular sieves to Metal Organic Frameworks) that would provide improved filtration of radon from xenon, and optimization of techniques to separate the two is important to reduce the challenge and risk of future xenon-based experiments [Sim].

Techniques for sensing radon in gases already have achieved sensitivities to 10 atoms, very close to the ultimate possible, using ultra-low background Lucas cells [Liu] or miniaturized proportional counters [Rau, ZS] and traps to extract the radon atoms from the gas. However, their slow response limits these techniques to $^{222}$Rn, and contamination of the counted sample is a concern. Development is needed to achieve similar sensitivity in chambers that can monitor the radon concentration continuously. Flowing air continuously past electrostatic detectors is the standard basic technique [Kiko], but detection efficiencies and backgrounds both likely could be improved. Robust, high-sensitivity detection of radon will permit better measurement of the radon emanation rate from materials used in detectors, provide better screening of some materials than is possible with traditional HPGe counters [9], and allow continuous monitoring of low-radon environments, in order to allow future generations of experiments to achieve lower backgrounds and better sensitivity.

Better mitigation against radon and radon-daughter backgrounds requires improvement of facilities including (1) reduced-radon storage capability, (2) reduced-radon laboratory spaces for detector assembly, (3) specialized radon filtration systems for liquids and gases used in detectors, (4) surface screeners sensitive to the non-penetrating radon daughters, and (5) methods of removing implanted radon daughters from surfaces.

## VI. Conclusions

A comparison of the planned requirements for Generation-2 dark matter experiments assay needs (Appendix B summarized in section IV) to the existing assay facilities identified for use by the underground physics research community (Appendix C) appears to show an inadequate level of sensitivity and through-put to achieve the background goals of the proposed experiments on a 3-4 year timeline expected for construction of Generation-2 dark matter experiments.

These needs will require investment in the tools needed to measure such radiopurity, both new techniques to improve sensitivity and increased throughput in moderate-sensitivity production screening. While mass spectrometry and atom trap techniques can be done above ground, radiometric assay techniques require shielded and low-background environments to obtain the required sensitivity. A community-wide assessment, as begun here, will help determine the level of additional screening capability required from the differing radiometric and non-radiometric assay methods.

Early investment in material assay capability can reduce the risk of later beginning an experimental construction project while the backgrounds from the materials remain an unknown. This early investment helps ensure the successful deployment of the next generation dark matter, neutrino, and neutrinoless double beta decay experiments.

The capability shortage is international, so sending samples overseas for assay is not the solution, although experiments with significant international collaborators will certainly employ sensitive screeners at European and Asian labs for a subset of samples and should be accounted for when determining the gap between assay need and current capability.

New models of collaborative work in assay and related infrastructure should be explored in order to find a sustainable way to address the shortfall and efficiently build up assay capacity.  A staged transition from an experiment-specific model to a multi-site user coordinated network would strengthen capabilities that have been built up over many years in the service of individual projects and capitalize on various individuals' specific expertise. Such an organization would also promote rapid and broad dissemination of research results.  When combined with access to assay and a database of previously assayed materials, this will significantly increase scientific reach and reduce risk for all rare event searches underground.

# References

[Facilities]. More details on facilities can be found at
https://zzz.physics.umn.edu/lowrad/consortium#available_assay_technology

[Survey]. Survey raw data is found at Detailed summary in CF1 paper and
http://snowmass2013.org/tiki-index.php?page=materials+details.

[Nico]. J. Nico, A. Piepke, and T. Shutt, NUSL White Paper: Ultra Low Background Counting Facility (Oct 2001, Lead, SD)

[Cushman]. P. Cushman, DUSEL S1 Technical Report (Dec 13, 2006): Low Level Counting Infrastructure and Requirements.
http://www.deepscience.org/TechnicalDocuments/Final/lowlevelcounting_final.pdf

[Ali]. G. Alimonti, et al. (Borexino-CTF Collaboration), Astropart. Phys., 8, 141 (1998)

[XIAa]. W. K. Warburton, J. Wahl, and M. Momayezi, "Ultra-low background gas-filled alpha counter," U.S. Patent 6 732 059, May 4, 2004.

[XIAb]. W. K. Warburton, B. Dwyer-McNally, M. Momayezi, and J. W. Wahl, *Ultra-low background alpha particle counter using pulse shape analysis*, IEEE Nuclear Sci. Symp. Conf. Rec., Oct. 16–22, 2004, vol. 1, pp. 577–581, Paper N16-80.

[Lawson] I. Lawson, *Low Background Counting At SNOLAB*, Presentation at the 2010 Low Radioactive Techniques conference. (August 28, 2010).

[Laub]. M. Laubenstein, *Overview of the screening activities with HPGe detectors*, Presentation at the 2013 Low Radioactive Techniques conference. (April 10-12, 2010).

[Heu]. G. Heusser, "GIOVE (Germanium spectrometer with Inner and Outer Veto), a new low background Ge-spectrometer at MPI-K," Presentation at the 2013 Low Radioactive Techniques conference. (April 10-12, 2010).

[HLN]. G. Heusser, M. Laubenstein, H. Neder, "Low-level germanium gamma-ray spectrometry at the μBq/kg level and future developments towards higher sensitivity", Radioactivity in the Environment, Volume 8, 2006, Pages 495-510.

[Xu]. Xu, J. et al, "A Study of the Residual 39Ar Content in Argon from Underground Sources," arXiv:1204.6011 (2012).

[Nac]. A. Nachab. Radon reduction and radon monitoring in the NEMO experiment. In P. Loaiza, editor, AIP Conf. Proc. 897: Topical Workshop on Low Radioactivity Techniques: LRT 2006, pages 35–39, Melville, NY, October 2007. American Institute of Physics.

[Simgen]. H. Simgen. Radon Assay and Purification Techniques. In L. Miramonti and L. Pandola, editors, Topical Workshop on Low Radioactivity Techniques: LRT 2013, AIP Conf. Proc. 1549, pages 102-107 (2013).

[Poc]. A. Pocar. Low background techniques for the Borexino nylon vessels. In B. Cleveland, R. Ford, and M. Chen, editors, Topical Workshop on Low Radioactivity Techniques: LRT 2004., volume 785 of American Institute of Physics Conference Series, pages 153–162, September 2005, arXiv:physics/0503243.

[Sch]. R. W. Schnee, R. Bunker, G. Ghulam, D. Jardin, M. Kos, and A. Tenney. Construction and mea- surements of a vacuum-swing-adsorption radon-mitigation system. In L. Miramonti and L. Pandola, editors, Topical Workshop on Low Radioactivity Techniques: LRT 2013, AIP Conf. Proc. 1549, pages 116-119 (2013).

[Gal]. C. Galbiati, private communication.

[Gra]. D. Grant, A. Hallin, S. Hanchurak, C. Krauss, S. Liu, and R. Soluk. Low Radon Cleanroom at the University of Alberta. In R. Ford, editor, Topical Workshop on Low Radioactivity Techniques: LRT 2010, volume 1338 of American Institute of Physics Conference Series, pages 161–163, April 2011.

[Heu]. G. Heusser, W Rau, B. Freudiger, M. Laubenstein, M. Balata, and T. Kirsten. 222Rn detection at the muBq/m3 range in nitrogen gas and a new Rn purification technique for liquid nitrogen. Applied Radiation and Isotopes, 52:691–695, 2000.

[SZ]. H. Simgen and G. Zuzel. Analysis of the 222rn concentration in argon and a purification technique for gaseous and liquid argon. Applied Radiation and Isotopes, 67(5):922–925, May 2009.

[Liu]. M. Liu, H. W. Lee, and A. B. McDonald. 222Rn emanation into vacuum. Nuclear Instruments and Methods in Physics Research A, 329:291–298, May 1993.

[Rau]. W. Rau and G. Heusser. 222Rn emanation measurements at extremely low activities. Applied Radiation and Isotopes, 53:371–375, 2000.

[ZS]. G. Zuzel and H. Simgen. High sensitivity radon emanation measurements. Applied Radiation and Isotopes, 67:889–893, 2009.

[Kiko]. J. Kiko. Detector for 222Rn measurements in air at the 1mBq/m3 level. Nuclear Instruments and Methods in Physics Research A, 460:272–277, March 2001.

[IEEE]. "IEEE Test Procedures for Germanium Detectors for Ionizing Radiation," ANSI/IEEE Standard 325-1986.

[Chen]. C.Y. Chen et al., "Ultrasensitive isotope trace analyses with a magneto-optical trap," Science 286 (Iss. 5442) 1139-1141 (1999).

[Collon]. P. Collon et al., "Tracing noble gas radionuclides in the environment," Ann. Rev. Nucl. Part. Sci. 54 39-67 (2004).

[Jeynes]. C. Jeynes et al., "Elemental thin film depth profiles by ion beam analysis using simulated annealing - a new tool," J. Phys. D-App. Phys. 36(7) R97-R126 APR 7 2003.

[Comp]. R.J. Composto et al., "Application of ion scattering techniques to characterize polymer surfaces and interfaces," Mat. Sci. Eng. R-Rep. 38(3-4) 107-180 JUL 1 2002.

[Reij]. J. Reijonen, et al., NIM A522 (2004) 598-602.

[Parry]. S.J. Parry, "Activation Spectrometry in Chemical Analysis". John Wiley and Sons: New York, NY. (1991)

[Gold]. T. Goldbrunner et al., J. Radioanal. Nuc. Chem. 216 (1997) 293.

[Djurcic]. Z. Djurcic et al., NIM A507 (2003) 680. (hep-ex/0210038).

# Appendix A – Background reported in low background experiments

This table is intended to be representative only and not all-inclusive. The one finding to draw from this table is: Most dark matter experiments are dominated by a background-contributing source that is within the experimental construction materials. Non-internal backgrounds *could have been* cosmic ray muon-induced backgrounds, $^{238}$U spontaneous fission neutrons from the underground cavern, U/Th decay chain induced ($\alpha$,n) reaction neutrons from the underground cavern, direct radon infiltration (as oppose to radon exposure during assembly), gamma-rays from the underground cavern, and so on.

| EXPERIMENT | DOMINANT BACKGROUND | CITATION |
|---|---|---|
| CDMS-II Ge CDMS-II Si | Misidentification of surface events with an "analysis between alpha-decay and surface-event rates provides evidence that $^{210}$Pb (a daughter-product of $^{222}$Rn) is a major component of our surface event background." | Science 26 March 2010: Vol. 327 no. 5973 pp. 1619-Science 26 March 2010: Vol. 327 no. 5973 pp. 1619-1621 and Phys. Rev. Lett. 102, 011301 (2009) 1621 and Phys. Rev. Lett. 102, 011301 (2009), arXiv:1304.3706, arXiv:1304.4279 |
| CoGeNT | Likely g-rays from U/Th/K in front end electronics (resistor) | Physical Review D 88, 012002 (2013) |
| COUPP 4-kg | U-238 spontaneous fission neutrons and U/Th decay chain induced ($\alpha$,n) in PZT piezoelectric transducers | PRD 86, 052001 (2012) |
| CRESST-II | Recoiling Pb-206 nuclei from Po-210 decays (likely radon progeny) | Eur. Phys. J. C (2012) 72:1971 |
| DAMIC SNOLAB | Version 1 of aluminum nitride substrate contained U-238 limiting sensitivity | 33$^{RD}$ INTERNATIONAL COSMIC RAY CONFERENCE, RIO DE JANEIRO 2013, THE ASTROPARTICLE PHYSICS CONFERENCE |
| DM-Ice | "Dominant background in DM-Ice17: $^{40}$K & $^{210}$Pb in the crystals" | DM-Ice presentation: SNOWMASS 2013: Cosmic Frontier Workshop March 6 - 8, 2013 |
| DRIFT-IId | "…Radon Progeny Recoil (RPR) events, DRIFT's only known background…" likely in the cathode wires. | Astroparticle Physics Volume 35, Issue 7, February 2012, Pages 397–401 |
| EDELWEISS-II | Residual fiducial gamma-ray background leakage | PRD 86, 051701(R) (2012) |
| KIMS | Surface alpha emitters | PRL 108, 181301 (2012) |
| PICASSO | Alpha background of uncertain origin. Potentially alpha emitters in the $C_4F_{10}$ (e.g. decay chains below Ra) | Phys. Lett. B711 (2012) 153-161 |
| SIMPLE | Background neutrons originate mainly from the glass detector containment and shield water. | Phys. Rev. Lett. 108, 201302 (2012) |
| TEXONO | Likely $\gamma$-rays from U/Th/K similar to CoGeNT | Phys. Rev. Lett. 110, 261301 (2013) |
| XENON100 | Electron recoil background estimate including Gaussian and anomalous events | PRL 109, 181301 (2012) |
| XMASS | Radon daughters on PMTs and $^{14}$C on PMT seals | K. Hiraide, AXION-WIMP 2012 |
| ZEPLIN-III | First science run: "The FSR sensitivity was limited by background originating from PMT γ-rays." Second science run: The "electron recoil" leakage events from gamma-rays from PMTs was leading background. | Physics Letters B Volume 709, Issues 1–2, 13 March 2012, Pages 14–20 |

# Appendix B – Survey of G2 experiment assay needs

This table summarizes the answers to the survey that we received from each experiment. To view original material sent by each experiment, click on the experiment name [Survey].

http://www.snowmass2013.org/tiki-index.php?page=materials+details

| Experiment | Materials | Techniques | Samples per year | Contaminant | Sensitivity needed | Energy range | Sample preparation | Special handling | Proprietary? |
|---|---|---|---|---|---|---|---|---|---|
| DarkSide | stainless steel, copper, brass, PTFE, kovar, kapton, cuflon, electronic components, TMB (liquid) | HPGe, ICPMS, NAA | 30 - 50 | all parts for the present deployment of the DS50 TPC made of materials described above | 0.1 mBq/kg – 1 Bq/kg | 40keV – 3MeV | precision cleaning and surface treatment, sample preparation | handling in radon-free clean room, after cleaning store samples in radon-proof bags | No |
| C4 | Front-end electronics (various materials: metals, plastics, resistors) | HPGe | Probably only 1 sample | U-238, Th-232, and K-40 | < 0.1 mBq/kg | Up to 2.614 MeV (Tl-208 line) | Cannot be exposed to mine air, hopefully handled in cleanroom | No | No |
| DM-Ice | Stainless steel | HPGe | 5 | U, Th, K, Co-60 | 1 mBq/kg (U); 1 mBq/kg (Th); 5 mBq/kg (K); 1 mBq/kg (Co-60) | 30 – 3000 keV | surface etching | N/A | No |
| DM-Ice | Copper | HPGe | 5 | U, Th, K, Co-60 | 1 mBq/kg (U); 1 mBq/kg (Th); 5 mBq/kg (K); 1 mBq/kg (Co-60) | 30 – 3000 keV | surface etching | N/A | No |
| DM-Ice | Na Poweder Marinelli beakers | HPGe | 8 | U, Th, K | 10 ppt (U); 10 ppt (Th); 100 ppb (K) | 30 – 3000 keV | Dry box or glovebox for handling hygroscopic powder | N/A | No |
| DM-Ice | Quartz, solid | HPGe | 20 | U, Th, K | < 5 ppb (U); < 5 ppb (Th); < 5 ppm (K) | 30 – 3000 keV | | N/A | No |
| DM-Ice | Cables, spooled | HPGe | 3 | U, Th, K, Co-60 | 1 mBq/kg (U); 1 mBq/kg (Th); 5 mBq/kg (K); 1 mBq/kg (Co-60) | 30 – 3000 keV | surface etching | N/A | No |

| Experiment | Material | Technique | # Samples | Isotopes | Sensitivity | Energy Range | Cleaning | Handling | Published |
|---|---|---|---|---|---|---|---|---|---|
| DM-Ice | PMTs | HPGe | 20 | U, Th, K | < 5 ppb (U); < 5 ppb (Th); < 5 ppm (K) | 30 – 3000 keV | | N/A | No |
| DM-Ice | Optical grease / gel | HPGe | 4 | U, Th, K | 100 ppt (U); 100 ppt (Th); 100 ppb (K) | 30 – 3000 keV | | N/A | No |
| DM-Ice | PTFE | HPGe | 3 | U, Th, K, Co-60 | 100 ppt (U); 100 ppt (Th); 100 ppb (K); 0.1 mBq/kg (Co-60) | 30 – 3000 keV | surface etching | N/A | No |
| DM-Ice | Misc small | HPGe | 10 | U, Th, K, Co-60 | | 30 – 3000 keV | possibly surface etching | N/A | No |
| XENON | Stainless Steel | HPGe, ICP-MS, NAA, Rn emanation analysis | 2 months | 238U - 230Th, 226Ra - 206Pb, 232Th - Ac, 228Th - 208Pb, K-40, Co-60, Cs-137 | 1 mBq/kg | 30 – 2700 keV | ultrasound cleaning with weak acid solution | handle with sterile glove | No |
| XENON | PTFE | HPGe, ICP-MS, NAA, alpha counting, Rn emanation analysis | 2 months | 238U - 230Th, 226Ra - 214Po, 210Pb - 206Pb, 232Th-228Ac, 228Th - 208Pb, K-40, Co-60, Cs-137 | 50 micro-Bq/kg | 30 – 2700 keV | ultrasound cleaning with ethanol | handle with sterile glove | No |
| XENON | Oxgen free copper | HPGe, ICP-MS, NAA, alpha counting, Rn emanation analysis | 2 months | 238U - 230Th, 226Ra - 214Po, 210Pb - 206Pb, 232Th-228Ac, 228Th - 208Pb, 40K, 60Co, 137Cs | 50 micro-Bq/kg | 30 – 2700 keV | ultrasound cleaning with ethanol | handle with sterile glove | No |
| XENON | Photomultipliers | HPGe, Rn emanation analysis | 1.5 months | 238U - 230Th, 226Ra - 206Pb, 232Th-228Ac, 228Th - 208Pb, 40K, 60Co, 137Cs | <1 mBq/PMT | 30 – 2700 keV | wipe with ethanol | handle with sterile glove | No |

| Experiment | Material | Methods | Sample size | Isotopes | Sensitivity | Energy range | Cleaning | Handling | Screened |
|---|---|---|---|---|---|---|---|---|---|
| [XENON](#) | Resistors/capacitors for PMT electronics | HPGe, ICP-MS, NAAA, Rn emanation analysis | 1 month | 238U - 230Th, 226Ra - 206Pb, 232Th-228Ac, 228Th - 208Pb, 40K, 60Co, 137Cs | 1 micro-Bq/piece | 30 – 2700 keV | ultrasound bath with ethanol | handle with sterile glove | No |
| [XENON](#) | Internal cables | HPGe, ICP-MS, NAAA, Rn emanation analysis | 1 month | 238U - 230Th, 226Ra - 206Pb, 232Th - 228Ac, 228Th - 208Pb, 40K, 60Co, 137Cs | 1 mBq/kg | 30 – 2700 keV | ultrasound bath with ethanol | handle with sterile glove | No |
| [LZ](#) | Teflon | HPGe, Beta, NAA, Rn emanation analysis | 15 | 238U, 232Th, 40K, 210Pb | 0.1 mBq/kg | | | | Yes |
| [LZ](#) | Ti, Ti Alloy | HPGe, ICP-MS, NAA | 30 | 238U, 232Th, 40K, 60Co, 137Cs | 0.1 mBq/kg | | | | Yes |
| [LZ](#) | PMT + base components | HPGe, ICP-MS, NAA | 50 | 238U, 232Th, 40K | 0.1 mBq/kg | | | | Yes |
| [LZ](#) | TPC grid components | HPGe, ICP-MS | 50 | 238U, 232Th, 40K | 10 mBq/kg | | | | Yes |
| [LZ](#) | TPC HV cable | HPGe, NAA, Rn emanation | 10 | 238U, 232Th, 40K | 1 mBq/kg | | | | Yes |
| [LZ](#) | Insulation | HPGe, Rn emanation | 5 | 238U, 232Th, 40K, 137Cs | 1 mBq/kg | | | | Yes |
| [LZ](#) | Acrylic | NAA | 10 | 238U, 232Th, 40K | 0.1 mBq/kg | | | | Yes |
| [LZ](#) | Stainless Steel | HPGe | 10 | 238U, 232Th, 40K, 60Co, 137Cs | 1 mBq/kg | | | | Yes |
| [LZ](#) | Scintillator | NAA | 10 | 238U, 232Th, 40K | 0.01 mBq/kg | | | | Yes |
| [LZ](#) | Gd fluors | NAA, HPGe | 10 | 238U, 232Th, 40K | 1 mBq/kg | | | | Yes |

| Experiment | Component | Assay Method | Samples | Isotopes | Sensitivity | Energy Range | Cleaning | Handling | Rn Emanation |
|---|---|---|---|---|---|---|---|---|---|
| LZ | purification/recirculation | HPGe, NAA, ICPMS, Rn emanation | 20 | 238U, 232Th, 40K | 0.1 mBq/kg | | | | Yes |
| SuperCDMS | copper | HPGe, ICPMS, XIA | 24 | 238U, 232Th, 40K, 60Co | 0.02 - 0.1 mBq/kg (0.003 alphas/cm2/day) | 30 - 3000 keV | cleaning with ethanol | handle with sterile glove | No |
| SuperCDMS | poly | HPGe | 12 | 238U, 232Th, 40K, 60Co | 0.1 - 0.05 mBq/kg | 30 - 3000 keV | cleaning with ethanol | handle with sterile glove | No |
| SuperCDMS | Pb | HPGe | 12 | 238U, 232Th, 40K, 60Co | 0.05 mBq/kg | 30 - 3000 keV | none | handle with gloves | No |
| SuperCDMS | ceramic | HPGe, ICPMS | 12 | 238U, 232Th, 40K, 60Co | 0.05 mBq/kg | 30 - 3000 keV | clean with ethanol | handle with gloves | No |
| SuperCDMS | PMTs/SiPMs | HPGe, ICPMS | 24 | 238U, 232Th, 40K, 60Co | 3 ppb U,Th 100 ppm K | 30 - 3000 keV | cleaning with ethanol | handle with sterile glove | No |
| SuperCDMS | various electrical components | HPGe, ICPMS | 24 | 238U, 232Th, 40K, 60Co | 0.05 mBq/kg | 30 - 3000 keV | cleaning with ethanol | handle with sterile glove | No |
| SuperCDMS | Cu/Cirlex stiffner board | HPGe, ICPMS | 12 | 238U, 232Th, 40K, 60Co | 0.05 mBq/kg | 30 - 3000 keV | cleaning with ethanol | handle with sterile glove | No |
| SuperCDMS | flex cables | HPGe, ICPMS | 12 | 238U, 232Th, 40K, 60Co | 0.05 mBq/kg | 30 - 3000 keV | cleaning with ethanol | handle with sterile glove | No |
| SuperCDMS | 4K connectors | HPGe, ICPMS | 12 | 238U, 232Th, 40K, 60Co | 0.05 mBq/kg | 30 - 3000 keV | cleaning with ethanol | handle with sterile glove | No |
| SuperCDMS | scintillator | HPGe | 12 | 238U, 232Th, 40K, 60Co | | 30 - 3000 keV | 0.001 - 0.01 mBq/kg | | No |
| COUPP | fused silica | ICPMS | 10 | alpha emitters | 0.01 mBq/kg | > 3000 keV | surface etching (if underground) | No | No |
| COUPP | fused silica surfaces | alpha counting | 10 | alpha emitters | 0.01 alphas/cm2/day | > 3000 keV | surface etching (if underground) | No | No |
| COUPP | stainless | HPGe, | 10 | 238U | 0.01 mBq/kg | HPGe countin | surface etching (if | No | No |

|  | steel | ICPMS |  |  |  | g | underground) |  |  |
|---|---|---|---|---|---|---|---|---|---|
| COUPP | propylene glycol | HPGe, ICPMS | 10 | alpha emitters | 0.01 - 1 mBq/kg | HPGe counting | surface etching (if underground) | No | No |
| COUPP | seals | HPGe, ICPMS | 10 | alpha emitters | 0.01 - 10 mBq/kg | HPGe counting | surface etching (if underground) | No | No |
| COUPP | connectors | HPGe, ICPMS | 10 | alpha emitters | 0.5 - 10 mBq/kg | HPGe counting | surface etching (if underground) | handle with sterile glove | No |
| COUPP | wires | HPGe, ICPMS | 10 | alpha emitters | 0.1 - 10 mBq/kg | HPGe counting | surface etching (if underground) | handle with sterile glove | No |
| COUPP | cameras | HPGe, ICPMS | 10 | alpha emitters | 1 - 10 mBq/kg | HPGe counting | surface etching (if underground) | handle with sterile glove | No |
| COUPP | water | HPGe, LCS | 10 | alpha emitters | 1 - 10 mBq/kg | HPGe counting | surface etching (if underground) | handle with sterile glove | No |

## Appendix C – Materials and assay centers

Table 1. US Assay Resources [Facilities]

| Facility | Depth (mwe) | Suite of detectors and technology |
|---|---|---|
| Berkeley LBCF | surface | 2 HPGe (1 with muon veto) managed by LBNL<br>100% use for others<br>NaI, $BF_3$ counting, Shielded R&D space |
| PNNL | surface | ICPMS: Dedicated instrument and clean room facilities for low bkgd assay<br>6 commercially shielded HPGe detectors<br>    planning for use by others<br>Two 7-detector HPGe arrays (each ~400-500% relative efficiency) – unshielded. |
| PNNL Shallow Underground Lab | 30 | Copper electroforming and clean machining.<br>14-crystal HPGe array, considering use for others.<br>Multiple commercial HPGe for various stakeholders. |
| Oroville (LBNL) | 530 | 1 HPGe managed by LBNL, 100% use for others<br>Large Shielded R&D space |
| Kimballton (KURF) | 1450 | 2 HPGe managed by UNC/TUNL. 50% use by others |
| Soudan | 2100 | 1 HPGe managed by CDMS, 10% use by others.<br>1 HPGe managed by Brown, dedicated to LUX/LZ<br>6000 $m^3$ lab lined with muon tracker +<br>    2 muon-correlated neutron detectors<br>Large R&D space with muon tag provided |
| Homestake (SURF) | 4300 | 1 HPGe managed by CUBED, priority to LZ, Majorana, other users by negotiation.<br>Electroforming and clean machining currently exclusive use by Majorana. |

Table 2. International Assay Resources [Facilities]

| Facility | Depth (mwe) | Suite of detectors and technology |
|---|---|---|
| HADES | | |
| Japan | Surface | 1 HPGe (with active veto) managed by KamLAND, currently 100%, but may consider use for others.<br>1 HePG managed by CANDLES in Osaka (sea level), currently 90%, with 10% use for others. |
| Kamioka Observatory, Japan | 2700 | Each experimental group has their own devices for screening and assay, but will consider use by others.<br>1 HPGe managed by KamLAND, currently 100%<br>1 HPGe under construction by KamLAND: 100%<br>1 HPGe under construction by CANDLES: 100%<br>3 HPGe (2 p-type, 1 n-type)100% SuperK and XMASS<br>Underground ICP-MS and API-MS<br>    managed by SuperK and XMASS (100%).<br>Many radon detectors to measure radon emanation of materials, managed by SuperK and XMASS (100%) |
| CanFranc (LSC) Spain | 2450 | 5 HPGe p-type 100% usage by LSC.<br>    Outside collaboration possible<br>2 HPGe p-type to be installed by end of summer 2013 |
| Boulby Mine England | | |
| STELLA at LNGS Gran Sasso Italy | 3800 | 10 HPGe operated by INFN as a user facility<br>1 HPGe with 100% usage by XENON and GERDA, (DARWIN in future), Radon mitigation underway |
| LSM (Modane) France | 4800 | 15 HPGe with 6 dedicated to material selection.<br>- 2 detectors, 100% usage by SuperNEMO<br>- 1 detector 100% usage by EDELWEISS<br>- 3 detectors 100% dedicated to Modane exp experiments installed in Modane<br>2 detectors may be available to others at level of 5-10% |
| SNOLAB Canada | 6010 | 1 PGT coax HPGe 54% usage by Canadian based experiments, 34% usage by US based experiments, 12% usage by SNOLAB<br>1 Canberra well HPGe , 100% by SNO+ and DEAP<br>11 Electrostatic Counters (alpha counters), 100% usage by EXO, in future SNO+, PICASSO and MiniCLEAN<br>8 Alpha-Beta counters, 100% usage by SNO+<br>   available for other experiments on request<br>1 Canberra coax HPGe (currently being refurbished)<br>The SNOLAB facilities are used by SNOLAB based experiments, but can be negotiated during down time |
| CJPL (JinPing) China | 6800 | 1 HPGe managed by PandaX, 100% for PandaX<br>1 HPGe managed by CDEX, 90% usage by CDEX<br>2 HPGe to be installed by end of 2013: ~ 70% CDEX, ~30% availability reserved for others. |

# Appendix D – Natural uranium and thorium decay chains

## U-238 Decay Series

| Isotope | half-life | gamma energies (KeV) |
|---|---|---|
| U238 | $4.468 \times 10^9$ years | ---- |
| Th234 | 24.1 days | 63.3 (4.47%)<br>92.38 (2.60%)<br>92.80 (2.56%) |
| Pa234m | 1.17 minutes | 765 (0.207%)<br>1001 (0.59%) |
| 99.8%     0.13%<br>              Pa234 | 6.75 hours | 100 (50%)<br>700 (24%)<br>900 (70%) |
| U234 | $2.47 \times 10^5$ years | 53.2 (0.123%) |
| Th230 | $8.0 \times 10^4$ years | 67.7 (0.373%) |
| Ra226 | 1602 years | 186.2 (3.50%) |
| Rn222 | 3.823 days | 510 (0.076%) |
| Pc218 | 3.05 minutes | ------ |
| 99.98%     0.02%<br>Pb214 | 26.8 minutes | 53.2 (1.1%)<br>242.0 (7.46%)<br>295.2 (19.2%)<br>351.9 (37.1%)<br>785.9 (1.09%) |
|              At218 | 2 seconds | ----- |
| Bi214 | 19.7 minutes | 609.3 (46.1%)<br>768.4 (4.89%)<br>806.2 (1.23%)<br>934.1 (3.16%)<br>1120.3 (15.0%)<br>1238.1 (5.92%)<br>1377.7 (4.02%)<br>1408.0 (2.48%)<br>1509.2 (2.19%)<br>1764.5 (15.9%) |
| 99.98%     0.02%<br>Po214 | 164 microsec | 799 (0.014%) |
|              Tl210 | 1.3 minutes | 296 (80%)<br>795 (100%)<br>1310 (21%) |
| Pb210 | 21 years | 46.5 (4.05%) |
| Bi210 | 5.01 days | ------ |
| Po210 | 138.4 days | 803 (0.0011%) |
| Pb206 | Stable | ------ |

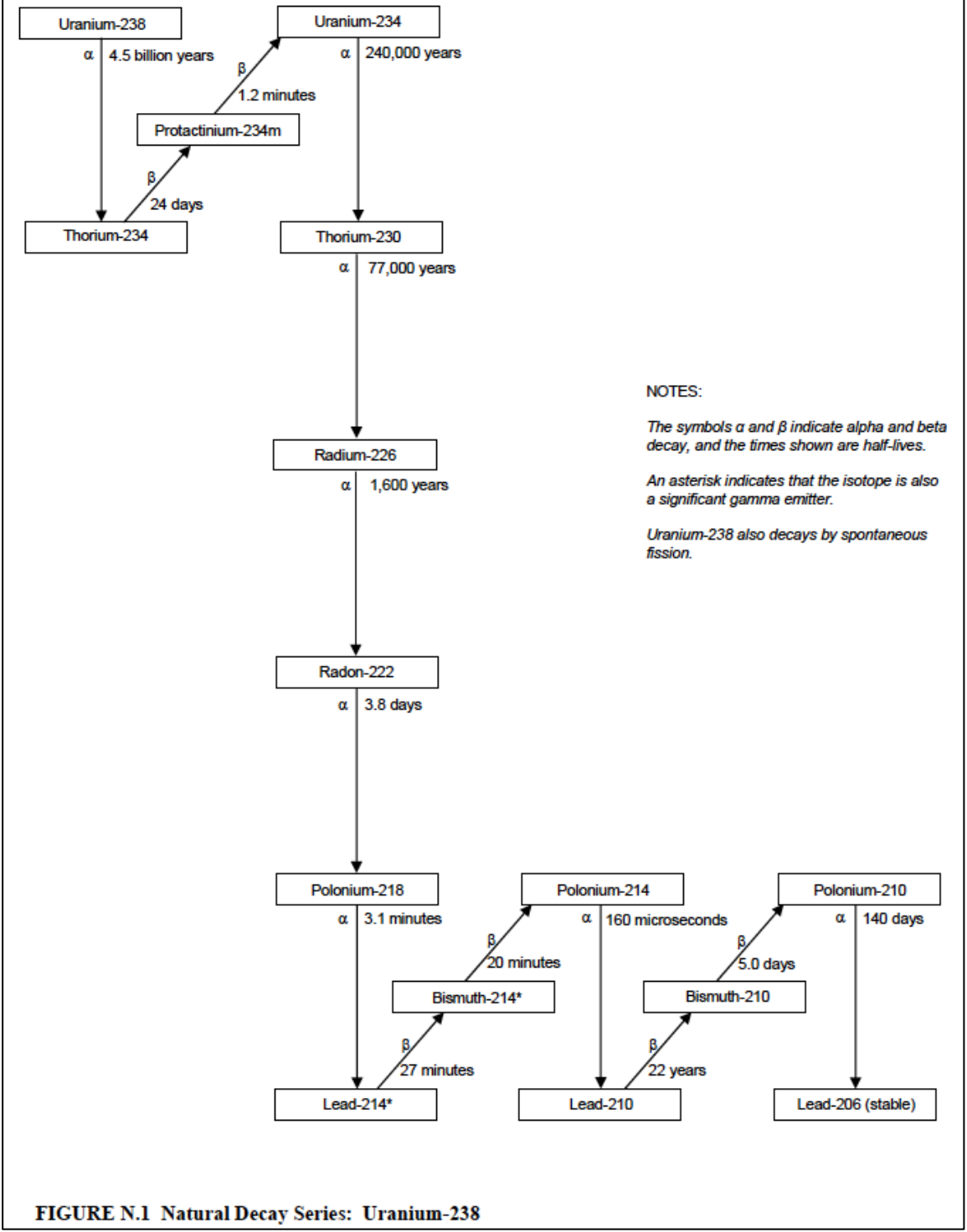

**FIGURE N.1 Natural Decay Series: Uranium-238**

## Th232 Decay Series

| Isotope | half-life | gamma energies (KeV) |
|---|---|---|
| Th232 | $1.405 \times 10^{10}$ years | 63.8 (0.267%) |
| Ra228 | 6.7 years | ------ |
| Ac228 | 6.13 hours | 57.7 (0.487%) |
| | | 99.5 (1.28%) |
| | | 129.0 (2.42%) |
| | | 154.0 (0.737%) |
| | | 209.3 (3.88%) |
| | | 270.2 (3.43%) |
| | | 328.0 (2.95%) |
| | | 338.3 (11.3%) |
| | | 409.5 (1.94%) |
| | | 463.0 (4.44%) |
| | | 772 (1.50%) |
| | | 794.9 (4.36%) |
| | | 835.7 (1.61%) |
| | | 911.2 (26.6%) |
| | | 964.8 (5.11%) |
| | | 969.0 (16.2%) |
| | | 1588.2 (3.27%) |
| Th228 | 1.91 years | 84.4 (1.22%) |
| Ra224 | 3.64 days | 241.0 (3.97%) |
| Rn220 | 55 seconds | 550 (0.07%) |
| Po216 | 0.15 seconds | ------ |
| Pb212 | 10.64 hours | 238.6 (43.6%) |
| | | 300.0 (3.34%) |
| Bi212 | 60.6 minutes | 39.9 (1.1%) |
| | | 727.3 (6.65%) |
| 64.06% Po212 / 35.94% Tl208 | 304 nsec / 3.1 minutes | ------- |
| | | 277.4 (6.31%) |
| | | 510.77 (22.6%) |
| | | 583.2 (84.5%) |
| | | 763.1 (1.81%) |
| | | 860.6 (12.4%) |
| | | 2614.5 (99%) |
| Pb208 | stable | ------- |

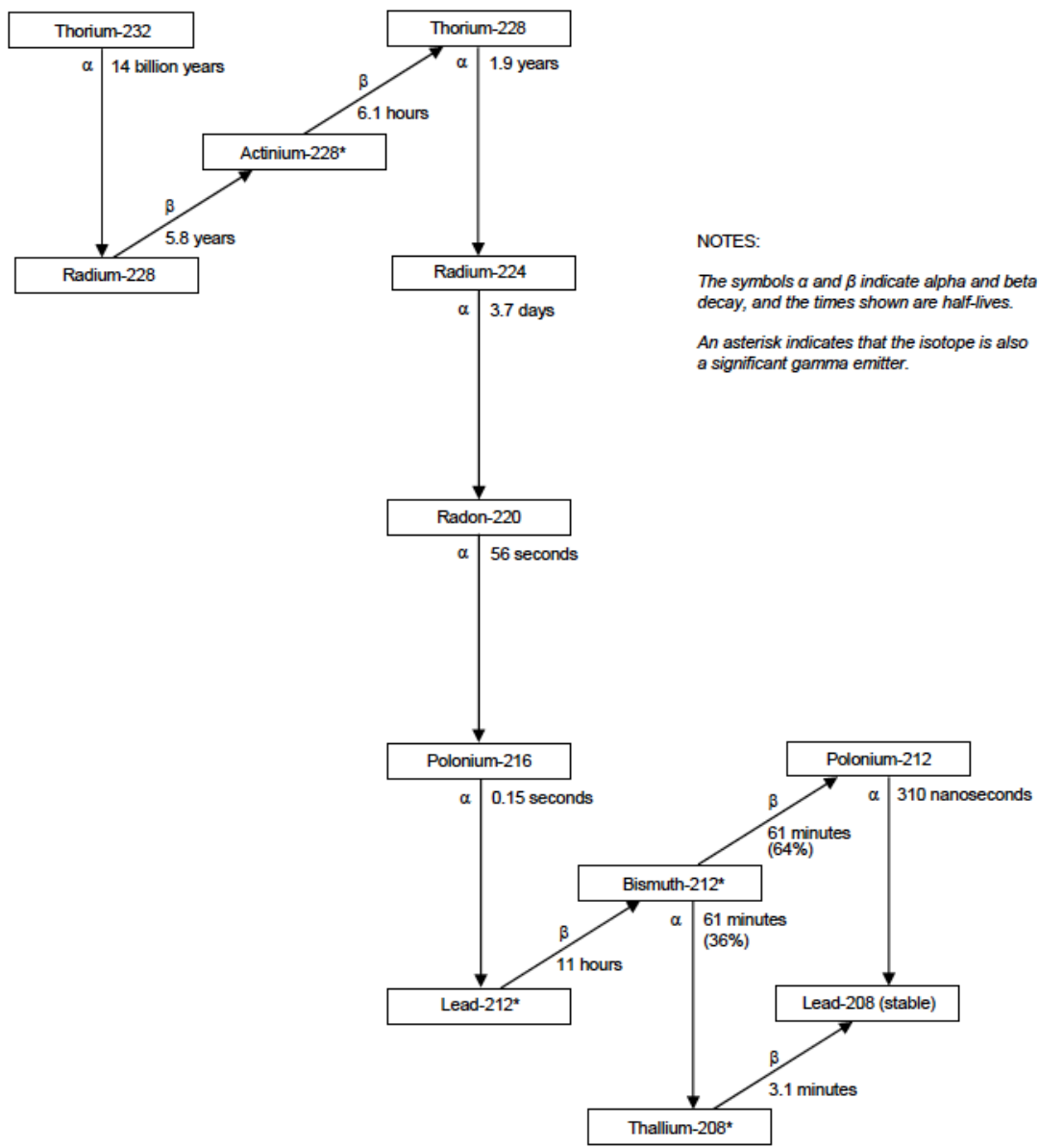

**FIGURE N.3 Natural Decay Series: Thorium-232**